\documentclass[graybox]{svmult}

\usepackage{type1cm}        
\usepackage{makeidx}         
\usepackage{graphicx}        
\usepackage{multicol}        
\usepackage[bottom]{footmisc}

\usepackage{newtxtext}       %
\usepackage[varvw]{newtxmath}       

\newcommand\be{\begin{eqnarray}}
\newcommand\ee{\end{eqnarray}}

\newcommand{\C}{{\mathbb C}}
\newcommand{\R}{{\mathbb R}}
\newcommand{\Z}{{\mathbb Z}}
\newcommand{\Oc}{{\mathbb O}}

\newcommand{\im}{{\rm i\,}}

\makeindex             

\begin{document}

\title*{Octonions, complex structures and Standard Model fermions}
\author{Kirill Krasnov\orcidID{0000-0003-2800-3767}}
\institute{Kirill Krasnov \at School of Mathematical Sciences, University of Nottingham, Nottingham, NG7 2RD, UK \\ \email{kirill.krasnov@nottingham.ac.uk}}
%
%
\maketitle

\abstract*{This article is a write-up of the talk given in one of the mini-symposia of the 2024 European Congress of Mathematicians. I will explain some basics of the representation theory underlying Spin(10) and SU(5) Grand Unified Theories. I will also explain the characterisation of the Standard Model gauge group $G_{\rm SM}$ as a subgroup of Spin(10) that was developed in \cite{Krasnov:2022meo}. Thus, the symmetry breaking required to obtain $G_{\rm SM}\subset {\rm Spin}(10)$ can be seen to rely on two suitably aligned commuting complex structures on $\R^{10}$. The required complex structures can in turn be encoded in a pair of pure spinors of ${\rm Spin}(10)$. The condition that the complex structures are commuting and suitably aligned translates into the requirement that the respective pure spinors are orthogonal and that their sum is again a pure spinor. The most efficient description of spinors, and in particular pure spinors of Spin(10) is via the octonionic model of the latter, and this is how octonions enter the story.}

\abstract{This article is a write-up of the talk given in one of the mini-symposia of the 2024 European Congress of Mathematicians. I will explain some basics of the representation theory underlying Spin(10) and SU(5) Grand Unified Theories. I will also explain the characterisation of the Standard Model gauge group $G_{\rm SM}$ as a subgroup of Spin(10) that was developed in \cite{Krasnov:2022meo}. Thus, the symmetry breaking required to obtain $G_{\rm SM}\subset {\rm Spin}(10)$ can be seen to rely on two suitably aligned commuting complex structures on $\R^{10}$. The required complex structures can in turn be encoded in a pair of pure spinors of ${\rm Spin}(10)$. The condition that the complex structures are commuting and suitably aligned translates into the requirement that the respective pure spinors are orthogonal and that their sum is again a pure spinor. The most efficient description of spinors, and in particular pure spinors of Spin(10) is via the octonionic model of the latter, and this is how octonions enter the story.}

\section{Introduction}
\label{sec:1}

This article is based on the material described in \cite{Krasnov:2022meo}. Our goal here is to describe the ideas behind the Spin(10) and SU(5) Grand Unified Theories of elementary particles from a more mathematical perspective. As such, our motivations are very similar to those of \cite{Baez:2009dj}. The motivational part of \cite{Baez:2009dj}, as well as some of the relevant representation theory is much more detailed than we attempt here, and the interested reader is advised to consult this reference for further details. The novelty of our presentation is that we will describe the symmetry breaking mechanism that produces the Standard Model gauge group 
\be
G_{\rm SM}={\rm SU}(3)\times{\rm SU}(2)\times{\rm U}(1)/\Z_6 \subset {\rm Spin}(10)
\ee
as the one that relies on two suitably aligned complex structures in $\R^{10}$. Also pure spinors and the octonionic model of Spin(10) are going to play an important role. 

We start, in Section \ref{sec:GUT}, with the description of the elementary particle content of the Standard Model (SM) of particle theory. As is was known since the 70's, i.e. almost immediately after the SM was formulated, this particle content arises very naturally if one embeds the SM gauge group $G_{\rm SM}$ into a larger gauge group such as SU(5) or Spin(10). Particle physicists usually refer to the later as SO(10). However, since spinors are going to play a very important role in our description of the symmetry breaking, we will refer to the relevant gauge group as Spin(10), to justify our usage of the spinor representations of the latter. 

Section \ref{sec:CS} is central to our presentation. Here we remind the reader the concept of the complex structure on a vector space, and also describe what happens when one has two commuting complex structures. We then describe the well-known relation between (complex lines of) pure spinors of ${\rm Spin}(2n)$ and complex structures on $\R^{2n}$. We end this section with Theorem \ref{thm} detailing the characterisation of $G_{\rm SM}\subset {\rm Spin}(10)$ that was developed in \cite{Krasnov:2022meo}.

Section \ref{sec:oct} describes the octonionic model of ${\rm Spin}(10)$, in which Weyl spinors become identified with 2-component columns with entries in complexified octonions $\Oc_\C$. This model allows for a very explicit description of pure spinors of ${\rm Spin}(10)$, detailed in Proposition \ref{prop}. We then explicitly describe the two pure spinors that are relevant for the symmetry breaking ${\rm Spin}(10)\to G_{\rm SM}$. We end with some discussion.

There is a large literature on the topics of this contribution. Some of the most relevant references are \cite{Gunaydin:1973rs}-\cite{Boyle:2020ctr}.

\section{Particle content of the SM, SU(5) and Spin(10) unification}
\label{sec:GUT}

\subsection{Particles}

Particles of the Standard Model come in three generations that are exact copies of each other as far as their $G_{\rm SM}$ transformation properties are concerned. We will only consider a single generation in this exposition. For readers without background in physics we remind that the factors of the SM gauge group $G_{\rm SM}$ have the following purpose. The SU(3) factor describes the strong force, which is the force that is responsible for keeping together the constituents (quarks) of particles such as proton and neutron (making up the atomic nuclei). The SU(2) factor describes the weak force that is behind nuclear reactions and the atomic bomb. Finally, the U(1) factor, or rather its certain combination with ${\rm U}(1)\subset {\rm SU}(2)$ describes the electromagnetic force. 

All particles of the SM, apart from their transformation properties with respect to $G_{\rm SM}$, also transform non-trivially with respect to the Lorentz group ${\rm Spin}(3,1)$. We recall that this group acts by pseudo-orthogonal transformations on Minkowski spacetime $\R^{3,1}$, which serves as the arena on which all particle physics takes place. In contrast, $G_{\rm SM}$ does not act on the spacetime $\R^{3,1}$. With respect to the Lorentz group ${\rm Spin}(3,1)$, all elementary particles transform as spinors. Recall that ${\rm Spin}(3,1)={\rm SL}(2,\C)$ has two types of (Weyl) spinor representations. Both types of spinors can be described as 2-component columns $\psi\in \C^2$ with complex entries. If we view an element $g\in {\rm SL}(2,\C)$ as a $2\times 2$ matrix with complex entries and unit determinant, then spinors of one type transform as $\psi\to g\psi$. The spinors of the other type transform as $\tilde{\psi}\to g^* \tilde{\psi}$, where $g^*$ is the complex conjugate matrix. It is clear that the 2-component spinors of type $\tilde{\psi}$ are complex conjugates of the spinors of type $\psi$. The field theory Lagrangian that encodes dynamics as well as interactions of the elementary particles is real, and involves one 2-component spinor for each particle, as well as the complex conjugate spinor. It is then convenient to describe all particles using the 2-component Lorentz spinors of the same type $\psi$, keeping it in mind that the complex conjugate spinors of the opposite type also appear in the Lagrangian. 

Having made the convention that we use 2-component Lorentz spinors of the same type to describe all particles, we obtain Table \ref{tab:1} of particles making up the first generation of the SM, together with their transformation properties with respect to $G_{\rm SM}$. The particles that are non-singlets with respect to the strong force ${\rm SU}(3)$, namely $Q, \bar{u}, \bar{d}$ are called quarks. The remaining particles $L, \bar{e}$ are called leptons. Two different 2-component spinors are needed to describe a particle with the same name (for all particles but the neutrino), e.g. both $u,\bar{u}$ are needed to describe the up quark {\bf and} its anti-particle. One can say that $u$ describes the up quark while $\bar{u}$ describes its anti-particle, although this terminology is empty unless one defines an independent notion of particles and anti-particles. We do not need this and will not attempt such a definition in this exposition. The other pairs are $d,\bar{d}$ that describe the down quark and its anti-particle, and $e,\bar{e}$ that describe the electron and its anti-particle (positron). The only 2-component spinor that does not have its partner is $\nu$, and this is because in the SM neutrino coincides with its anti-particle. However, the embedding of $G_{\rm SM}$ into ${\rm Spin}(10)$ predicts $\bar{\nu}$ as well, see below.

\begin{table}[!t]
\caption{Particles of one generation of the SM. The electric charge $Q$ of every particle arises as the combination of its $Y$-charge and the charge it has with respect to ${\rm U}(1)\subset{\rm SU}(2)$.}
\label{tab:1}       
%
%
\begin{center}
\begin{tabular}{ |c|c|c| c|c|c|} 
 \hline
 Particles & SU(3) & SU(2)  & $Y$ & $T^3$ & $Q=T^3+Y$ \\ 
 \hline
 $Q=\left( \begin{array}{c} u \\ d\end{array}\right)$ & $\bar{\C}^3$ & $\C^2$ & 1/6  & $\begin{array}{c} 1/2 \\ -1/2\end{array}$ & $\begin{array}{c} 2/3 \\ -1/3\end{array}$ \\
 $\begin{array}{c} \bar{u} \\ \bar{d}\end{array}$ & $\C^3$ & singlet & $\begin{array}{c} -2/3 \\ 1/3\end{array}$  & 0 & $\begin{array}{c} -2/3 \\ 1/3\end{array}$ \\
 $L=\left( \begin{array}{c} \nu \\ e \end{array}\right)$ & singlet & $\C^2$ & -1/2  & $\begin{array}{c} 1/2 \\ -1/2\end{array}$ & $\begin{array}{c} 0 \\ -1\end{array}$ \\
$ \bar{e}$ & singlet & singlet & 1 & 0 & 1 \\
 \hline
\end{tabular}
\end{center}
\end{table}

\subsection{Georgi-Glashow ${\rm SU}(5)$ GUT}

The Georgi-Glashow ${\rm SU}(5)$ GUT \cite{Georgi:1974sy} arises from the realisation that one can embed the rank 4 $G_{\rm SM}$ into the rank 4 simple Lie group ${\rm SU}(5)$. The embedding is as follows
\be\label{23-embedding}
 {\rm SU}(3)\times{\rm SU}(2)\times{\rm U}(1)_Y \ni (h_3, h_2, \alpha) \to \left( \begin{array}{cc} \alpha^2 h_3 & 0 \\ 0 & \alpha^{-3} h_2 \end{array}\right)\in {\rm SU}(5).
 \ee
 We can then attempt to identify the $\C^5$ representation of ${\rm SU}(5)$ with particles. It is clear that as the representation of ${\rm SU}(3)\times{\rm SU}(2)$ $\C^5$ will split into $\C^3\oplus\C^2$. It is also clear that the one will have $3Y_{\C^3} + 2Y_{\C^2}=0$. The only particle doublet that could be identified with $\C^2$ is $L$, which has the $Y$ charge of $-1/2$. The other particle then must have the $Y$ charge of $1/3$, and thus be identified with $\bar{d}$. All in all, the requirement that $\C^5$ of ${\rm SU}(5)$ can be identified with particles under the embedding of the type (\ref{23-embedding}) fixes the embedding to be
 \be\label{23-embedding*}
 {\rm SU}(3)\times{\rm SU}(2)\times{\rm U}(1)_Y \ni (h_3, h_2, e^{\im\phi}) \to \left( \begin{array}{cc} e^{\im\phi/3} h_3 & 0 \\ 0 & e^{-\im\phi/2} h_2 \end{array}\right)\in {\rm SU}(5),
 \ee
 with the convention that the 5-component columns that such matrices act on are in $\C^5$. The numbers appearing in the exponent of the ${\rm U}(1)$ factors on the diagonal are the $Y$-charges of the particles. 
 
 The other particles are identified as follows. We consider the representation $\Lambda^3 \C^5$. It splits as
 \be
  \Lambda^3(\C^3 \oplus \C^2)=  \Lambda^3 \C^3 \oplus (\Lambda^2 \C^3 \otimes \C^2) \oplus (\C^3\otimes \Lambda^2 \C^2). 
  \ee
  The first factor here has the $Y$-charge of $+1$ and is an ${\rm SU}(3)$ singlet. It must be identified with the partner $\bar{e}$ of the electron. The second factor transforms as $\Lambda^2 \C^3\sim \bar{C}^3$ under ${\rm SU}(3)$ and is an ${\rm SU}(2)$ doublet of $Y$-charge $2/3-1/2=1/6$. It must be identified with $Q$. The third factor is an ${\rm SU}(3)$ triplet with $Y$-charge $1/3-1=-2/3$, which is $\bar{u}$. This finishes the assignment of particles to representations of ${\rm SU}(5)$. Notably, all the particles of one SM generation fit perfectly into two irreducible representations of ${\rm SU}(5)$, namely $\C^5, \Lambda^3\C^5$. One certainly experiences a sense of magic in seeing this happen.  
  
Having considered $\Lambda^1 \C^5, \Lambda^3 \C^5$, it is very natural to consider also the representation $\Lambda^5 \C^5$. It splits as
 \be
 \Lambda^5 \C^5 = \Lambda^3 \C^3 \otimes \Lambda^2 \C^2.
 \ee
 Both factors are ${\rm SU}(3)$ and ${\rm SU}(2)$ singlets, and the total $Y$ charge is zero. This representation can be used to describe the partner $\bar{\nu}$ of the neutrino that is missing from the above table but is very natural from the point of view of ${\rm Spin}(10)$ GUT. 
 
 The preceding discussion can be summarised as follows. The particles of the Table \ref{tab:1} together with $\bar{\nu}$ become identified with the following representations of ${\rm SU}(5)$
 \be\label{split-GG}
 \Lambda^1 C^5 \oplus \Lambda^3 \C^5 \oplus \Lambda^5 \C^5 = \\ \nonumber
 (\C^3\oplus \C^2) \oplus (\Lambda^3 \C^3 \oplus \Lambda^2 \C^3 \otimes \C^2 \oplus \C^3\otimes \Lambda^2 \C^2) \oplus (\Lambda^3 \C^3 \otimes \Lambda^2 \C^2)= \\ \nonumber
 (\bar{d}, L) + (\bar{e}, Q, \bar{u}) + (\bar{\nu}).
 \ee
 The vector space on the left-hand side is the space of odd degree differential forms on $\C^5$. This space is also one of the two Weyl spinor representations of ${\rm Spin}(10)$, of which ${\rm SU}(5)$ is naturally a subgroup. So, while the particles of one generation of the SM are only partially unified by several different representations of ${\rm SU}(5)$, they become fully unified into a single irreducible representation of ${\rm Spin}(10)$ GUT. 
 
 \subsection{${\rm Spin}(10)$ GUT}
 
 It is a basic fact of the ${\rm Spin}(2n)$ representation theory that the unitary group ${\rm U}(n)$ is naturally a subgroup ${\rm U}(n)\subset {\rm Spin}(2n)$, and the under this embedding the spinor representation $S=\C^{2^n}$ splits as
 \be\label{spinor-un-split}
 S = \oplus_{k=0}^n \Lambda^k \C^n.
 \ee
 Further, the spinor representation of ${\rm Spin}(2n)$ splits into the sum of two Weyl representations, each of dimension $\C^{2^{n-1}}$, and under ${\rm U}(n)$ these transform as sums of even and odd degree terms in (\ref{spinor-un-split}). Applying this to the case of ${\rm Spin}(10)$ we get
 \be
 S_- = \Lambda^{odd}\C^5=\Lambda^1 \C^5 \oplus \Lambda^3 \C^5 \oplus \Lambda^5 \C^5,
 \ee
where $S_-$ on the left-hand side is one of the two chiral Weyl spinor representations of ${\rm Spin}(10)$. With our conventions $\Lambda^{even}\C^5$ would be identified with the other Weyl representation $S_+$. As we observed in the previous subsection, this is exactly the particle content of one general of the SM, plus $\bar{\nu}$ "sterile" neutrino that transforms in the trivial representation of all the factors of $G_{\rm SM}$. This gives a very compelling unification pattern. 
 
 ${\rm Spin}(10)$ unification traces its history to \cite{Fritzsch:1974nn}. While SU(5) unification is ruled out experimentally, at least in its simplest version, Spin(10) GUT is still an active field of study, as is exemplified by a recent paper \cite{Preda:2025afo}, see also references therein. The difficulty of this unification scenario is in the fact that an intricate pattern of symmetry breaking is required to go from Spin(10) to the gauge group that is unbroken in Nature, namely ${\rm SU}(3)\times{\rm U}(1)$. This requires a set of Higgs fields that effect the symmetry breaking. At least four such Higgs fields is necessary to build a realistic model, if one also only allows the renormalisable interactions, see \cite{Preda:2025afo} for a discussion. The arising model is quite baroque, there are many terms in the arising Lagrangian, and the difficulty of the Spin(10) unification is in giving a predictive model. In this exposition, we will describe a new characterisation of $G_{\rm SM}\subset {\rm Spin}(10)$, which points in the direction of Spin(10) model different from the ones currently explored.

\section{Complex structures and pure spinors}
\label{sec:CS}

We now recall some basics about (orthogonal) complex structures on $\R^{2n}$ and their relation with pure spinors of ${\rm Spin}(2n)$. 

\subsection{Complex structures}

Let $\langle\cdot,\cdot\rangle$ be the standard inner product on the Euclidean space $\R^{2n}$, and let $J:\R^{2n}\to \R^{2n}$ be an endomorphism that squares to minus the identity $J^2=-\mathbb{I}$ and also preserves the inner product $\langle J\cdot,J\cdot\rangle=\langle\cdot,\cdot\rangle$. Such a map $J$ is called an orthogonal complex structure. The eigenvalues of $J$ are $\pm \im$, and there are no real eigenspaces in $\R^{2n}$. However, the complexification $\R^{2n}_\C$ splits into two eigenspaces of $J$. Moreover, every real vector $v\in \R^{2n}$ can be uniquely represented as a sum of an eigenvector of $J$ and its complex conjugate, so we can write
\be
\R^{2n} = E_J \oplus \bar{E}_J, \qquad E_J=\{ v\in \R^{2n}_\C: Jv =\im v\} = \C^n.
\ee
This identifies $\R^{2n}\sim E_J=\C^n$ and thus gives $\R^{2n}$ the structure of a complex vector space. It is then a standard fact that the subgroup of ${\rm SO}(2n)$ that commutes with $J$ is ${\rm U}(n)$. Indeed, orthogonal transformations that commute with $J$ leave $E_J$ invariant, and thus can only act on it by ${\rm U}(n)$ transformations. 

Let us now consider a somewhat less standard setup of two commuting orthogonal complex structures
\be
J_1, J_2: J_1^2=J_2^2=-\mathbb{I}, \qquad [J_1,J_2]=0.
\ee
It is easy to see that the product $K=J_1 J_2$ is an orthogonal transformation that squares to plus the identity $K^2=\mathbb{I}$. The eigenvalues of $K$ are $\pm 1$, and the action of $K$ further splits both $E_{J_1}, E_{J_2}$. We have
\be
E_{J_1} = E_{J_1}^+ \oplus E_{J_1}^-,
\ee
where $E_{J_1}^-\subset E_{J_2}$ is the part of $E_{J_1}$ that is shared with $E_{J_2}$, and $E_{J_1}^+\subset \bar{E}_{J_2}$ is the part of $E_{J_1}$ that is shared with $\bar{E}_{J_2}$. It is similarly easy to see that the subgroup of ${\rm SO}(2n)$ that commutes with both $J_1, J_2$ is ${\rm U}(k)\times{\rm U}(n-k)$, where $k$ is the (complex) dimension of $E_{J_1}^+$. 

For the case of interest ${\rm SO}(10)$, the two possible subgroups arising as those commuting with a pair of complex structures $J_1, J_2: [J_1, J_2]=0$ are
\be\label{scenarios}
{\rm U}(4)\times {\rm U}(1) \qquad \text{and} \qquad {\rm U}(3)\times {\rm U}(2).
\ee
It is the second of this that is relevant to the SM, but both are important if we want to break the symmetry from ${\rm Spin}(10)$ to the group unbroken in Nature, namely ${\rm SU}(3)\times{\rm U}(1)$. To obtain the characterisation of $G_{\rm SM}$ that we are after we need the notion of pure spinors. 

\subsection{Pure spinors}

Pure spinors of ${\rm Spin}(2n)$ are Weyl spinors in an orbit of this group that has the smallest dimension (thus largest stabiliser). The stabiliser of a pure spinor is ${\rm SU}(n)$. There is a one-to-one correspondence between projective pure spinors of ${\rm Spin}(2n)$ of given parity and orthogonal complex structures on $\R^{2n}$ inducing a given orientation. By projective pure spinors we mean pure spinors $\psi$ modulo multiplication by a non-zero complex number $\psi\sim \lambda\psi, \lambda\in \C^*$. 

A pure spinor $\psi$ of ${\rm Spin}(2n)$ thus endows $\R^{2n}$ with a complex structure $J_\psi$ and identifies $\R^{2n}\sim E_{J_\psi}=\C^n$. It also endows $\C^n$ with a top form in $\Lambda^n \C^n$, and this is why the stabiliser of $\psi$ is only ${\rm SU}(n)$ rather than ${\rm U}(n)$. 

Pure spinors of ${\rm Spin}(10)$ can be characterised very concretely. First, we note that 
\be
{\rm dim}({\rm Spin}(10)/{\rm SU}(5)) = 45-24=21.
\ee
This is the dimension of the space of unit spinors $\psi\in \C^{16}$ satisfying 5 complex constraints. It is well-known that the constraints guaranteeing that $\psi$ is pure are quadratic in $\psi$, see e.g. formula (103) in \cite{Krasnov:2022meo} for their explicit characterisation, and also below. 

\subsection{New characterisation of $G_{\rm SM}$}

With these preparations in place, we can state the new characterisation of the SM gauge group described in \cite{Krasnov:2022meo}:
\begin{theorem} \label{thm} Let $\psi_1,\psi_2 \in S_-$ be two pure spinors of ${\rm Spin}(10)$ that have the properties: (i) spinors $\psi_{1,2}$ are orthogonal $\langle \hat{\psi}_1,\psi_2\rangle=0$. Here $\langle S_+,S_-\rangle$ is the ${\rm Spin}(10)$-invariant pairing and $\,\hat{} :S_-\to S_+$ is the ${\rm Spin}(10)$-invariant complex conjugation that maps Weyl spinors of one chirality into those of the other chirality; (ii) The sum $\psi_1+\psi_2$ is still a pure spinor. Then the subgroup of ${\rm Spin}(10)$ that stabilises one of the pure spinors, and projectively stabilises the other is $G_{\rm SM}$. 
\end{theorem}

We now sketch the proof of this statement. The key point is that both pure spinors $\psi_{1,2}$ give rise to complex structures $J_{\psi_1}, J_{\psi_2}$. When the spinors are orthogonal $\langle \hat{\psi}_1,\psi_2\rangle=0$ these complex structures commute. These means that there are just two possible scenarios (\ref{scenarios}) for the stabiliser of both of the complex structures. One observes that the situation when $\psi_1+\psi_2$ is still a pure spinor corresponds to the second scenario. Imposing the further condition that one of the pure spinors is stabilised rather than projectively stabilised selects the subgroup of ${\rm U}(3)\times{\rm U}(2)\subset {\rm SU}(5)$, which is precisely $G_{\rm SM}$. 

\section{Octonionic model}
\label{sec:oct}

All the considerations above can be made completely explicit by introducing an octonionic model of ${\rm Spin}(10)$. The reason for the existence of this model is that real Weyl spinors of ${\rm Spin}(8)$ {\bf are} octonions. This means that ${\rm Spin}(10)$ that contains ${\rm Spin}(8)$ as a subgroup also admits an octonionic description. 

\subsection{Octonionic model of ${\rm Spin}(10)$}

Concretely, let us consider the following $4\times 4$ matrices with entries in ${\rm End}(\Oc)$
\be\label{gamma}
\Gamma^x = \left( \begin{array}{cccc} 0 & 0 & 0 & L_{\bar{x}} \\ 0 & 0 & L_x & 0 \\ 0 & L_{\bar{x}} & 0 & 0 \\ L_x & 0 & 0 & 0 \end{array}\right), \quad
\Gamma^9 = \left( \begin{array}{cccc} 0 & 0 & 1 & 0 \\ 0 & 0 & 0 & -1 \\ 1 & 0 & 0 & 0 \\ 0 & -1 & 0 & 0 \end{array}\right), \quad 
\Gamma^{10} = \left( \begin{array}{cccc} 0 & 0 & -\im & 0 \\ 0 & 0 & 0 & -\im \\ \im & 0 & 0 & 0 \\ 0 & \im & 0 & 0 \end{array}\right).
\ee
Here $L_x$ is the operator of left multiplication by an octonion $x\in \Oc$, and $\bar{x}$ is the conjugate octonion. It is easy to check that these matrices generate the Clifford algebra $\Gamma^I \Gamma^J +\Gamma^J \Gamma^I=2\delta^{IJ} \mathbb{I}, I,J=1,\ldots,10$. Their commutators thus generate the Lie algebra $\mathfrak{so}(10)$.

The commutators of the $\Gamma$-matrices (\ref{gamma}) are block-diagonal. This means that the Lie algebra $\mathfrak{so}(10)$, in one of its two semi-spinor representations, can be represented by $2\times 2$ matrices with complexified octonionic entries:
\be\label{spin-10} 
\left( \begin{array}{cc} A+ \im a & - L_{\bar{x}}+\im L_{\bar{y}} \\ L_x +\im L_y  & A'-\im a \end{array}\right), \qquad A,A' \in \mathfrak{so}(8), \quad x,y\in \Oc, \quad a\in \R.
\ee
This gives an explicit description of the Lie algebra ${\mathfrak so}(10)$ in terms of $2\times 2$ matrices with complexified octonionic entries, acting on 2-columns with entries in complexified octonions. This gives the octonionic model of the $\mathfrak{so}(10)$ algebra, and also identifies each of the Weyl spinor representations of ${\rm Spin}(10)$ with $\Oc^2_\C$, i.e. 2-component columns with complexified octonions as entries. The described octonionic model of ${\rm Spin}(10)$ is also contained in \cite{Bryant}, with some differences in conventions. 

\subsection{Explicit description of pure spinors of ${\rm Spin}(10)$}

In the described octonionic model, every Weyl spinor of ${\rm Spin}(10)$ can be written as a 2-component column
\be
\psi = \left( \begin{array}{c} \alpha_1 +\im \alpha_2 \\ \beta_1+\im \beta_2\end{array}\right).
\ee
Here $\alpha_{1,2},\beta_{1,2}\in \Oc$ are octonions and $\im=\sqrt{-1}$. We have the following ${\rm Spin}(10)$-invariant Hermitian inner product on the space of Weyl spinors
\be
\langle \hat{\psi},\psi\rangle = |\alpha_1|^2 + |\alpha_2|^2 + |\beta_1|^2 + |\beta_2|^2.
\ee
We then have the following
\begin{proposition}\label{prop}
A Weyl spinor $\psi$ of ${\rm Spin}(10)$ is pure if and only if the following conditions are satisfied
\be
(\alpha,\alpha)=(\beta,\beta)=0, \qquad \alpha\cdot \bar{\beta}=0.
\ee
\end{proposition}
Here $(\cdot,\cdot)$ is the inner product in $\Oc$. Given two real octonions $\alpha,\beta\in \Oc$ we have $(\alpha,\beta)=\alpha\cdot\bar{\beta}$, where dot denotes the octonionic product. The inner product is extended to $\Oc_\C$ by linearity, so that $(\alpha,\alpha) = |\alpha_1|^2 - |\alpha_2|^2 + 2\im (\alpha_1,\alpha_2)$. In the last condition $\alpha\cdot \bar{\beta}$ is the octonionic product, and $\bar{\beta} = \bar{\beta}_1+\im \bar{\beta}_2$, so the bar operation acts only on the real octonions and does not touch the imaginary unit. 

It can be shown, see \cite{Krasnov:2022meo}, that a complexified null octonion, such as $\alpha,\beta$ in the above proposition, describes a pure spinor of ${\rm Spin}(8)$. This means that a pure spinor of ${\rm Spin}(10)$ is a pair of pure spinors of ${\rm Spin}(8)$ that are orthogonal in an appropriate sense. 

\subsection{Explicit description of the spinors relevant to ${\rm Spin}(10)\to G_{\rm SM}$}

We now describe explicitly a pair of pure spinors of ${\rm Spin}(10)$ that effect the symmetry breaking to $G_{\rm SM}$. These are given by
\be
\psi_1 =\left(\begin{array}{c} 1+\im {\bf u} \\ 0 \end{array}\right), \qquad \psi_2 =\left(\begin{array}{c} 0 \\ 1+\im {\bf u} \end{array}\right).
\ee
Here ${\bf u}$ is a unit imaginary octonion, and $1$ is the octonionic unit. Both spinors $\psi_{1,2}$ are pure because $1+\im {\bf u}$ is a complexified null octonion. The spinors $\psi_{1,2}$ are orthogonal. Indeed, the ${\rm Spin}(10)$-invariant inner product is given by
\be
\langle \hat{\psi}, \tilde{\psi}\rangle = (\alpha_1,\tilde{\alpha}_1) +  (\alpha_2,\tilde{\alpha}_2) + (\beta_1,\tilde{\beta}_1) + (\beta_2,\tilde{\beta}_2).
\ee
Here $\alpha_{1,2},\beta_{1,2}$ are the real octonions components of $\psi$, and $\tilde{\alpha}_{1,2},\tilde{\beta}_{1,2}$ are components of $\tilde{\psi}$. The sum of two spinors $\psi_{1,2}$ is
\be
\psi=\psi_1+\psi_2 = \left(\begin{array}{c} 1+\im {\bf u} \\ 1+\im {\bf u}  \end{array}\right).
\ee
This is a pure spinor because both of its complexified octonion entries are null $\alpha=\beta=1+\im {\bf u}$, and $(1+\im {\bf u})\cdot \overline{(1+\im {\bf u})}=  (1+\im {\bf u})\cdot(1-\im {\bf u}) = 0$. It can be explicitly checked, for example using the description (\ref{spin-10}) of the Lie algebra $\mathfrak{so}(10)$, that the subalgebra of $\mathfrak{so}(10)$ that stabilises $\psi_1$ and projectively stabilises $\psi_2$ is the Lie algebra of $G_{\rm SM}$. This is done in Section 7.3 of \cite{Krasnov:2022meo}.

It is interesting to remark that in the language of octonions the choice that is required to effect the breaking of symmetry to $G_{\rm SM}$ is that of a unit imaginary octonion ${\bf u}$. 
Such a choice of a unit imaginary octonion is also central to the scenario described in \cite{Dubois-Violette:2016kzx}, \cite{Todorov:2018mwd}, \cite{Dubois-Violette:2018wgs}.

\section{Discussion}

We end with a short discussion as to what the facts described may entail for particle physics. Our key observation is that it is possible to effect the symmetry breaking ${\rm Spin}(10)\to G_{\rm SM}$ by two Higgs fields both in the (Weyl) spinor representation of ${\rm Spin}(10)$. However, the symmetry that is observed as unbroken in Nature is even smaller than $G_{\rm SM}$, and is ${\rm SU}(3)\times{\rm U}(1)$. The SM of particle physics has a dedicated field -- the Higgs field -- that breaks $G_{\rm SM}\to {\rm SU}(3)\times{\rm U}(1)$. It is not difficult to see that our symmetry breaking scenario by pure spinors and associated complex structures can be generalised to effect also this symmetry breaking. For this one just needs to add yet another commuting complex structure $J_3$ that together with $J_1$ follow the first pattern in (\ref{scenarios}). The corresponding pure spinor is 
\be
\psi_3 = \left(\begin{array}{c} 1-\im {\bf u} \\ 0 \end{array}\right).
\ee
Taken together, the two pure spinors $\psi_1,\psi_3$ and their associated complex structures do not lead to a realistic symmetry breaking pattern. But supplemented with $\psi_2$ and its complex structure the symmetry breaking is precisely down to ${\rm SU}(3)\times{\rm U}(1)$. Again, this can be verified by an explicit Lie algebra computation with the octonionic model. 

Further, for aesthetic reasons it may be desirable to work not with 3 but with 4 Higgs fields all in the spinor representation. One then observes that adding
\be
\psi_4 = \left(\begin{array}{c}  0\\ 1-\im {\bf u} \end{array}\right)
\ee
does not destroy the symmetry breaking pattern that leads to ${\rm SU}(3)\times{\rm U}(1)$, with the condition that $\psi_{1,3}$ are stabilised while $\psi_{2,4}$ are only projectively stabilised. We also remark that the described pure spinors $\psi_{1,2,3,4}$ can all be aligned with particles. Indeed, all particles are identified with certain vectors in the Weyl spinor representation of ${\rm Spin}(10)$, and so $\psi_{1,2,3,4}$ each points in the direction of some particle. Using the identification described in \cite{Krasnov:2022meo}, it is not difficult to see that 
\be
\psi_1 \sim \bar{\nu}, \qquad \psi_2\sim \bar{e}, \qquad \psi_3\sim \nu, \qquad \psi_4\sim e.
\ee

The above discussion suggests that it could be very interesting to construct a ${\rm Spin}(10)$ GUT model that has 4 Higgs fields, all in the Weyl spinor representation of ${\rm Spin}(10)$. As far as we know, no models with only spinor Higgs fields was ever studied. We hope that this exposition together with work \cite{Krasnov:2022meo} will motivate construction and study of such a model in the future.

\begin{acknowledgement}
The author is grateful to the organisers of the ECM9 mini-symposium "Generalisations of complex analysis and its applications" for the invitation to present his research, and to the symposium participants for a stimulating environment. 
\end{acknowledgement}

\end{document}